\documentclass[3p,times,procedia]{elsarticle}

\usepackage{ecrc}

\volume{00}

\firstpage{1}

\journalname{Physics Procedia}
\jnltitlelogo{Physics Procedia}

\runauth{K. Rielage et al.}

\jid{phpro}

\CopyrightLine{2014}{Published by Elsevier Ltd.}

\usepackage{amssymb}

\begin{document}

\begin{frontmatter}




\title{Update on the MiniCLEAN Dark Matter Experiment}


\author[label3]{K. Rielage}
\author[label3]{M. Akashi-Ronquest}
\author[label6]{M. Bodmer}
\author[label3]{R. Bourque}
\author[label4]{B. Buck}
\author[label9]{A. Butcher}
\author[label8]{T. Caldwell}
\author[label12]{Y. Chen}
\author[label5]{K. Coakley}
\author[label3]{E. Flores}
\author[label4]{J.A. Formaggio}
\author[label1]{D. Gastler}
\author[label6]{F. Giuliani}
\author[label6]{M. Gold}
\author[label9]{E. Grace}
\author[label3]{J. Griego}
\author[label4]{N. Guerrero}
\author[label11]{V. Guiseppe}
\author[label7]{R. Henning}
\author[label3]{A. Hime}
\author[label4]{S. Jaditz}
\author[label1]{C. Kachulis}
\author[label1]{E. Kearns}
\author[label4]{J. Kelsey}
\author[label8]{J.R. Klein}
\author[label8]{A. Latorre}
\author[label10]{I. Lawson}
\author[label1]{S. Linden}
\author[label3]{F. Lopez}
\author[label13]{D.N. McKinsey}
\author[label7]{S. MacMullin}
\author[label8]{A. Mastbaum}
\author[label11]{D.-M. Mei}
\author[label9]{J. Monroe}
\author[label9]{J.A. Nikkel}
\author[label3]{J. Oertel}
\author[label2]{G.D. Orebi Gann}
\author[label4,label10]{K. Palladino}
\author[label11]{G. Perumpilly}
\author[label3]{L. Rodriguez}
\author[label12]{R. Schnee}
\author[label8]{S. Seibert}
\author[label9]{J. Walding}
\author[label12]{B. Wang}
\author[label6]{J. Wang}
\author[label12]{C. Zhang}

\address[label1]{Department of Physics, Boston University, Boston, MA 02215 USA}
\address[label2]{Department of Physics, University of California, Berkeley, Berkeley, CA 94720 USA}
\address[label3]{Los Alamos National Laboratory, Los Alamos, NM  87545 USA}
\address[label4]{Department of Physics, Massachusetts Institute of Technology, Cambridge, MA 02139 USA}
\address[label5]{National Institute of Standards and Technology, Boulder, CO 80305 USA}
\address[label6]{University of New Mexico, Albuquerque, NM 87131 USA}
\address[label7]{Department of Physics and Astronomy, University of North Carolina, Chapel Hill, NC 27599 USA}
\address[label8]{Department of Physics and Astronomy, University of Pennsylvania, Philadelphia, PA 19104 USA}
\address[label9]{Department of Physics, Royal Holloway, University of London, Egham, Surrey TW20 0EX UK}
\address[label10]{SNOLAB Institute, Lively, ON P3Y 1N2 Canada}
\address[label11]{Department of Physics, University of South Dakota, Vermillion, SD 57069 USA}
\address[label12]{Physics Department, Syracuse University, Syracuse, NY 13244 USA}
\address[label13]{Department of Physics, Yale University, New Haven, CT 06520 USA}

\begin{abstract}
The direct search for dark matter is entering a period of increased sensitivity to the hypothetical Weakly Interacting Massive Particle (WIMP).  One such technology that is being examined is a scintillation only noble liquid experiment, MiniCLEAN.  MiniCLEAN utilizes over 500 kg of liquid cryogen to detect nuclear recoils from WIMP dark matter and serves as a demonstration for a future detector of order 50 to 100 tonnes.  The liquid cryogen is interchangeable between argon and neon to study the A$^{2}$ dependence of the potential signal and examine backgrounds.  MiniCLEAN utilizes a unique modular design with spherical geometry to maximize the light yield using cold photomultiplier tubes in a single-phase detector. Pulse shape discrimination techniques are used to separate nuclear recoil signals from electron recoil backgrounds.  MiniCLEAN will be spiked with additional $^{39}$Ar to demonstrate the effective reach of the pulse shape discrimination capability.  Assembly of the experiment is underway at SNOLAB and an update on the project is given.
\end{abstract}


\end{frontmatter}


\section{Introduction}
Recently, the search for Weakly Interacting Massive Particle (WIMP) dark matter has seen the entrance of a number of experiments that utilize liquid noble elements (i.e. Xe, Ar, and Ne).  The MiniCLEAN experiment exploits the scintillation light from interactions in its target volume of liquid argon or neon to discriminate between electron recoil events and nuclear recoil events \cite{BH, clean}.  Referred to as a 'single-phase' noble liquid experiment because it only utilizes the liquid phase of the target, MiniCLEAN relies on pulse shape discrimination to reject background events that arise from gammas and betas.  This discrimination is essential when using liquid argon as the WIMP target as natural argon contains about one part in $10^{15}$ of $^{39}$Ar, a $\beta$-emitter with a half-life of 269 years.  To separate these beta events from a possible signal in natural argon, pulse shape discrimination capability is needed at the level of $10^{-9}$ or better.  This discrimination can be achieved by the unique nature of the scintillation light in liquid noble gases.  The scintillation light is produced from two dimer states: a singlet state with a short lifetime and a triplet state with a longer lifetime.  In argon and neon, the difference between the two lifetimes is long enough (ns vs. $\mu$s) that each scintillation event is composed of prompt light (the first $\sim$100 ns) and late light (the next 10-15 $\mu$s).  Furthermore, electronic recoil events in the liquid produce more triplet state dimers (and thus late light) than nuclear recoil events.  By using the ratio of prompt to total scintillation light one can discriminate between these types of events.   Small prototype experiments \cite{Lipp, Nik, DEAP1} have shown that this discrimination technique can reach the necessary capability.  The discrimination requirement in liquid neon is much lower since there are no natural occurring $\beta$-emitters in neon.  By building a detector capable of interchanging liquid argon and neon as targets one can examine any potential signal and external backgrounds by taking advantage of the expected A$^2$ dependence of WIMP interactions.  Furthermore, a large multi-ton single-phase detector filled with liquid neon would be sensitive to \emph{pp} solar neutrinos as well as dark matter \cite{clean, argon}.  

Backgrounds in the experiment originate from a variety of sources.  Internal gammas and neutrons from the detector materials are minimized by the choice of low radioactivity materials and by the use of shielding with additional cryogen and acrylic.  In addition, by reconstructing the position of the event in the detector, a central region of the target volume can be chosen (a fiducial volume) where few background events can interact.   A fiducial volume of approximately 150 kg is expected for MiniCLEAN but will depend upon the final position reconstruction uncertainties.  For surface background events originating from radon daughter plate out, the sensitive components of MiniCLEAN have been assembled in a strictly controlled environment and care will be taken to remove possible radon contamination during circulation and purification of the cryogen.  Finally, for backgrounds outside the detector, a water shield tank surrounds the detector with an active muon veto.

The MiniCLEAN experiment has three primary goals.  The MiniCLEAN detector will serve as a technical proof-of-principle and demonstrate all of the salient features of a $4\pi$ single-phase detector using, interchangeably, targets of LAr and LNe.  The experiment will also develop a robust analysis program where all detector parameters and response to signal and backgrounds are over-constrained through simulation and calibration.  Finally, MiniCLEAN will serve as a prototype to a full-scale, multi-ton CLEAN experiment.  This article presents a brief overview of the MiniCLEAN detector, its current status, and recently developed analysis techniques.

\section{The MiniCLEAN Detector}
The MiniCLEAN detector is a conceptually simple detector.  A target volume of $\sim$500 kg of LAr or LNe is contained within a sphere coated with wavelength shifter to convert the ultraviolet scintillation light (128 nm for argon and 80 nm for neon) to the visible.  Light guides bring the visible light to photomultiplier tubes where the signal is recorded.  By using the scintillation light distribution and pulse shape discrimination, the position, energy, and type of event are reconstructed.  

\begin{figure}[htbp]
\begin{center}
\includegraphics[width=5in]{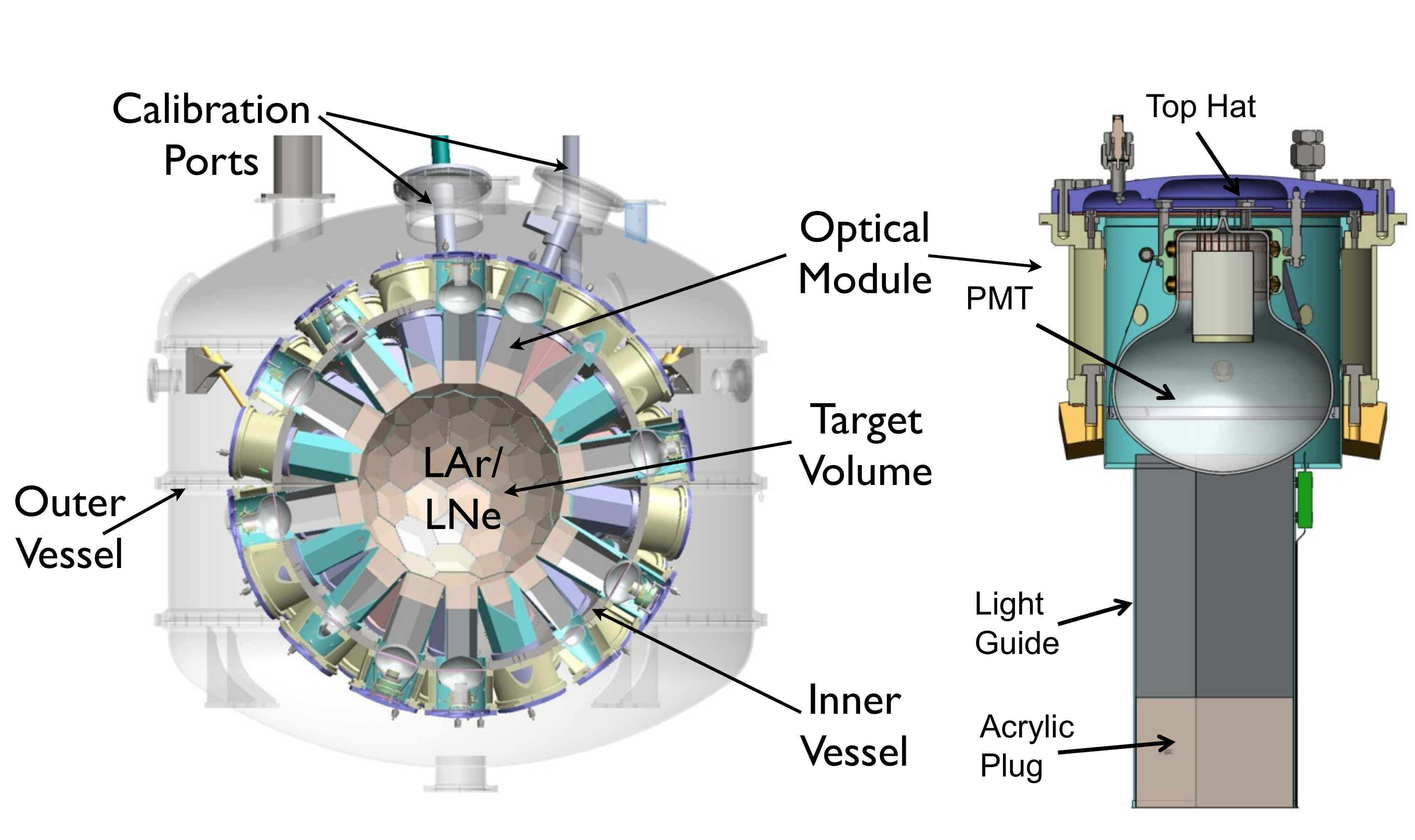}
\caption{Model of the MiniCLEAN detector showing the optical modules.}
\label{model}
\end{center}
\end{figure}

\begin{figure}[htbp]
\begin{center}
\includegraphics[width=5in]{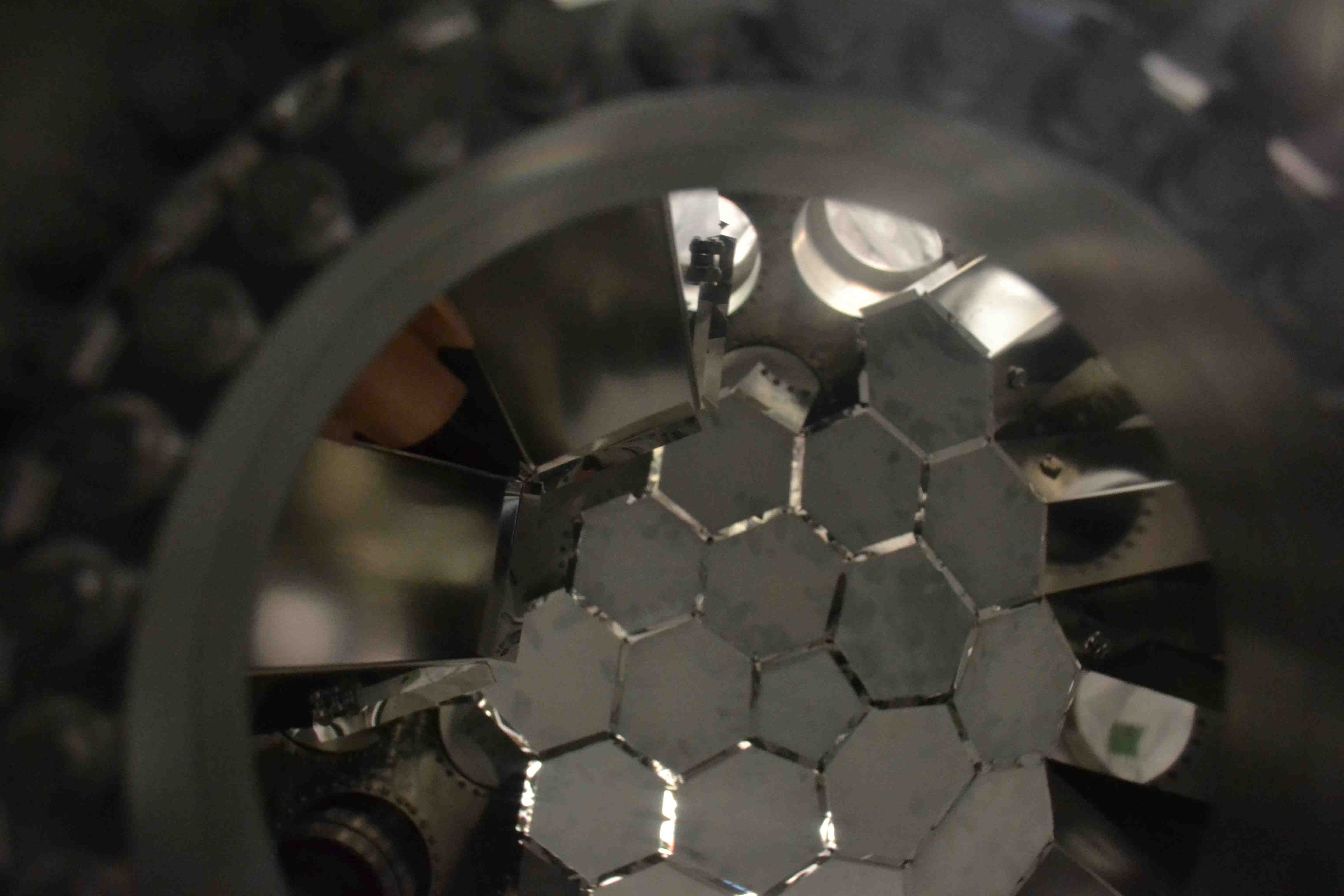}
\caption{View into the Inner Vessel before the final optical module was installed showing the acrylic plug front surfaces.}
\label{inside}
\end{center}
\end{figure}

The MiniCLEAN detector is undergoing finally assembly and commissioning at the SNOLAB facility in Sudbury, Canada.  The MiniCLEAN detector is shown in Figure \ref{model}.  The central detector is composed of a stainless steel Inner Vessel that contains the liquid cryogen and 92 optical cassette modules.  Each optical module contains an 8'' Hamamatsu R5912-02Mod photomultiplier tube capable of operating in cryogen \cite {pmt} in a stainless steel light guide with an acrylic plug at the end.  The inside of the light guides are lined with Vikuiti\texttrademark ESR foil by 3M \footnote{Certain commercial equipment, instruments, or materials are identified in this paper to foster understanding. Such identification does not imply recommendation or endorsement by the National Institute of Standards and Technology, nor does it imply that the materials
or equipment identified are necessarily the best available for the purpose.} to improve their reflectivity and the foil continues past the front of the light guide to fill in the gaps between modules.  The acrylic plugs are shaped like hexagons and pentagons and their front surfaces form a 92-sided sphere that contains the cryogenic volume.  The acrylic was made by Spartech \cite{acrylic}.  The front surface of the acrylic plugs are coated with a thin layer of tetraphenyl butadiene \cite{TPB}, a wavelength shifter.   Figure \ref{inside} shows the front of some of the acrylic plugs inside the Inner Vessel before the final assembly was completed.  Each optical module was assembled in a reduced radon environment underground.  They were installed into the 92 ports on the Inner Vessel while under a flow of nitrogen gas from liquid nitrogen boil off.  This assembly sequence greatly reduced the expected surface backgrounds on the modules from radon daughter plate out.  

\begin{figure}[htbp]
\begin{center}
\includegraphics[width=5in]{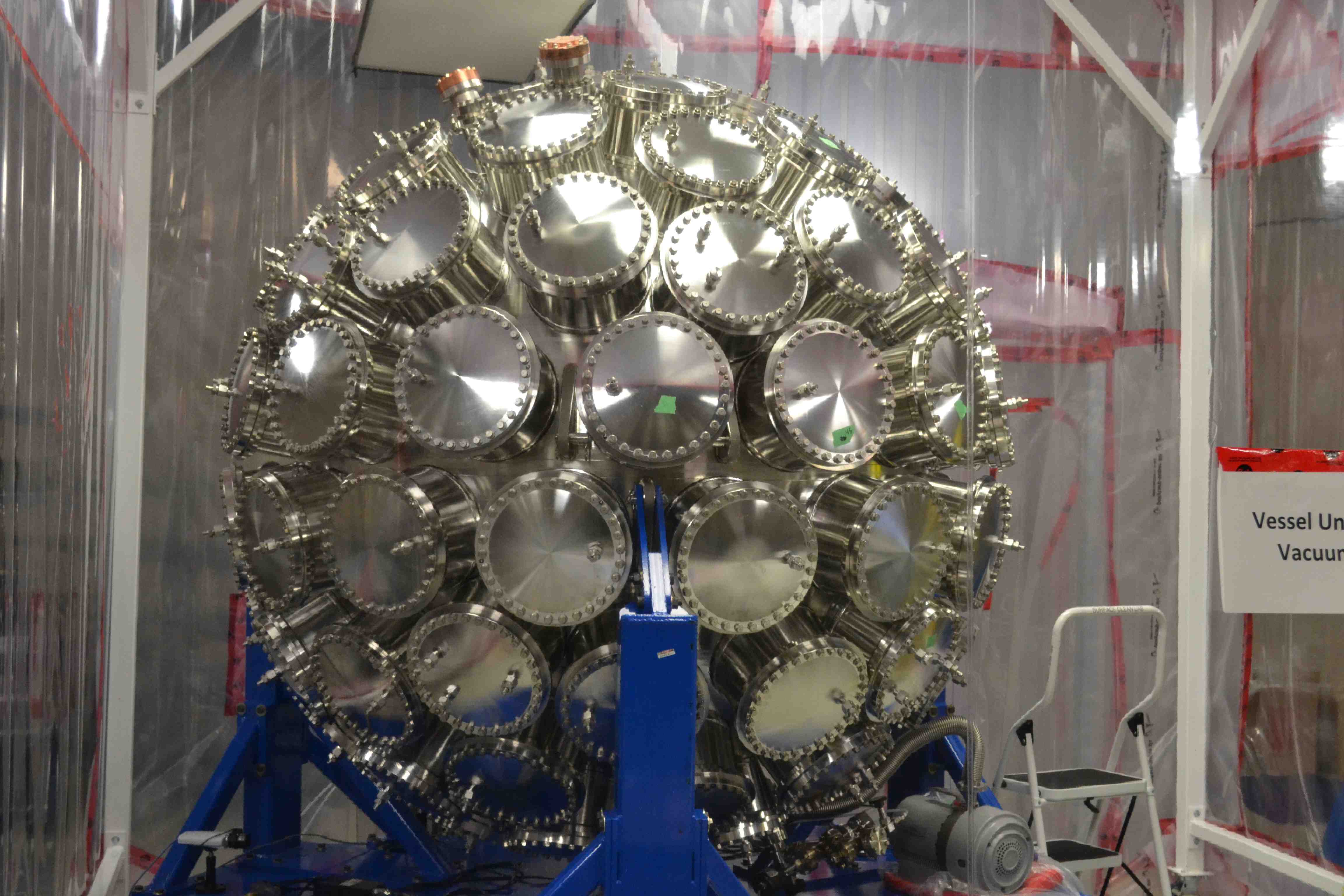}
\caption{Fully assembled MiniCLEAN Inner Vessel in a softwall cleanroom underground at SNOLAB.}
\label{IV}
\end{center}
\end{figure}

\begin{figure}[htbp]
\begin{center}
\includegraphics[width=2.5in]{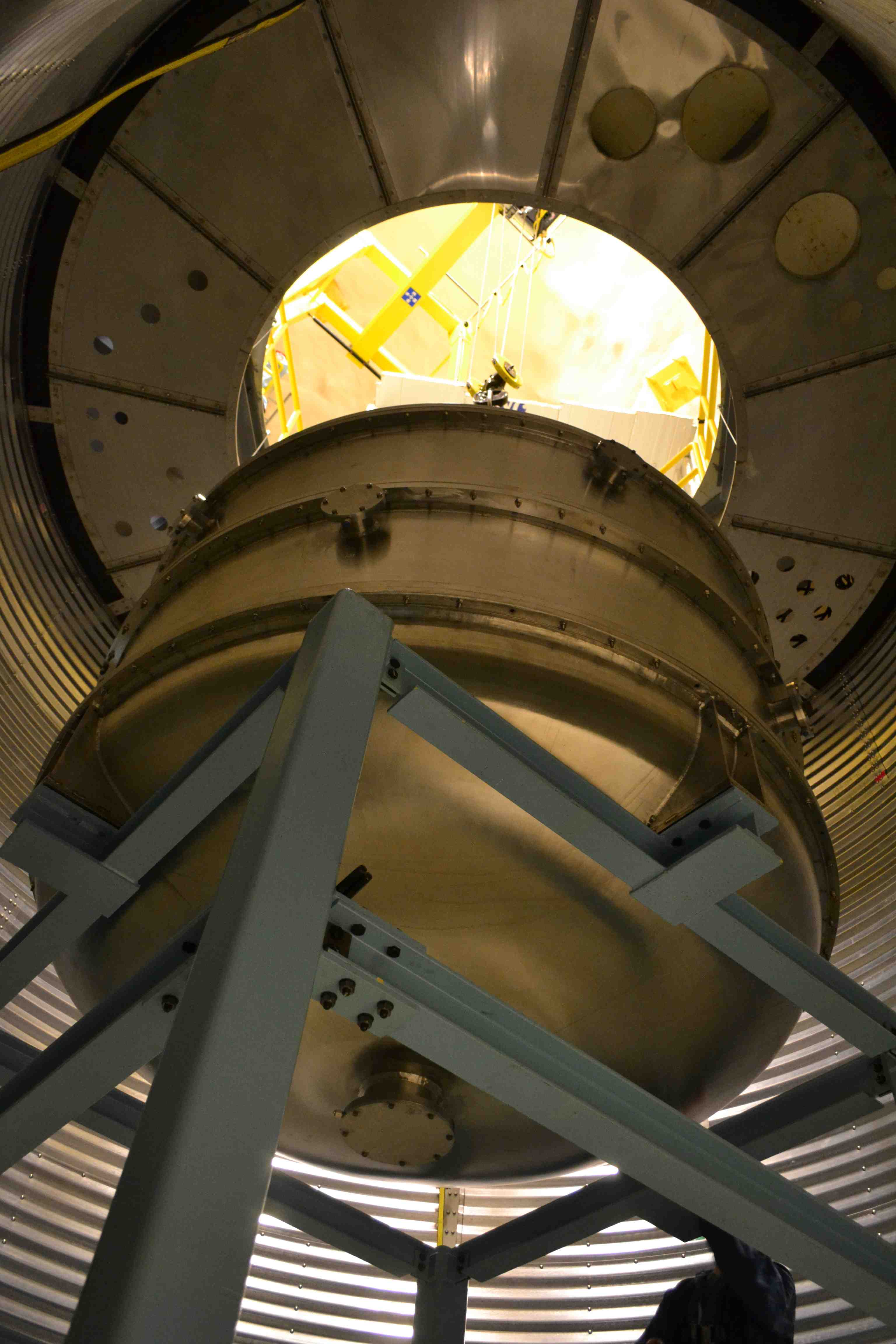}
\caption{The MiniCLEAN Outer Vessel on top of the detector stand inside the water shield tank in the SNOLAB Cube Hall.}
\label{OV}
\end{center}
\end{figure}

\begin{figure}[htbp]
\begin{center}
\includegraphics[width=5in]{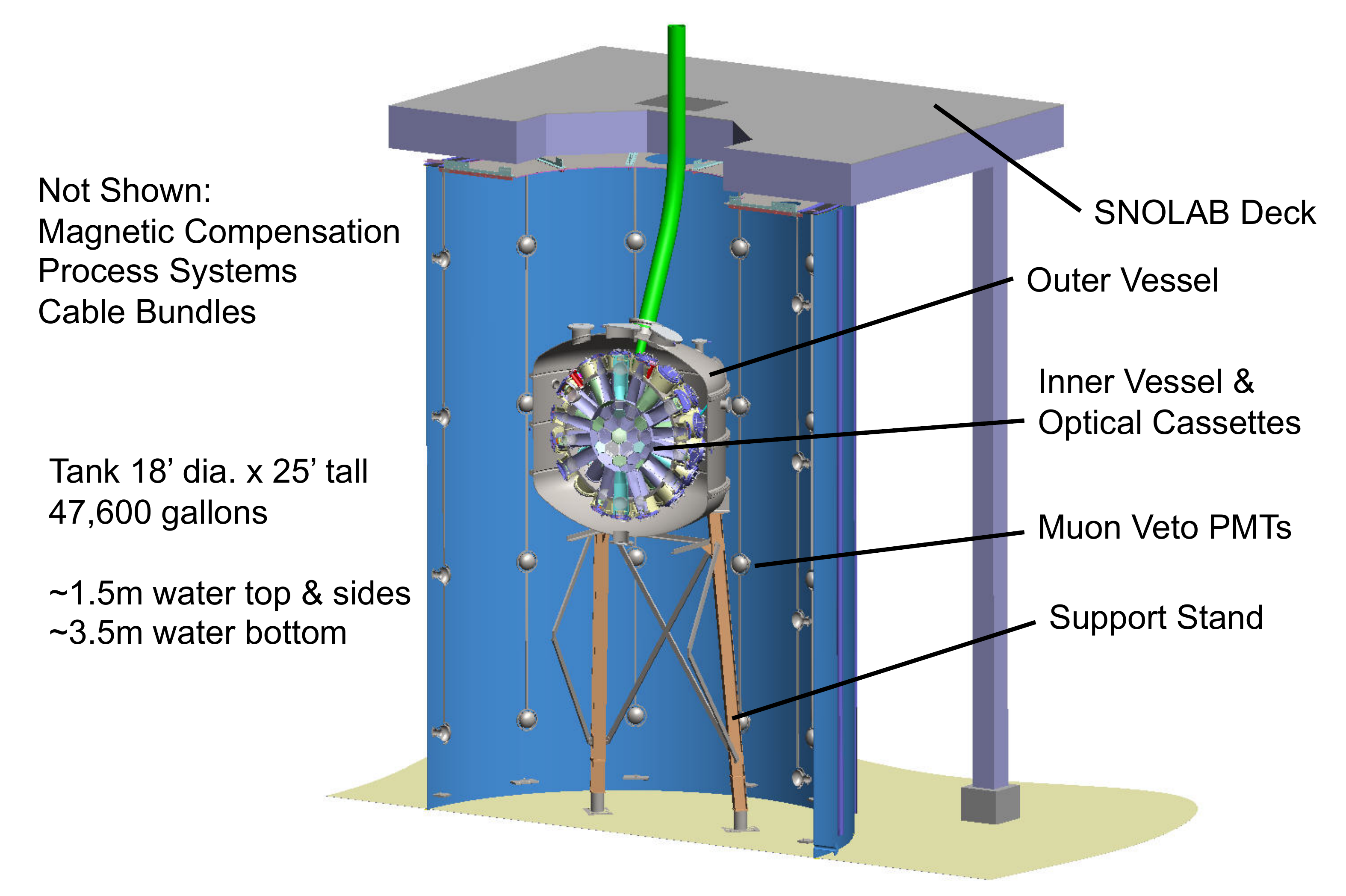}
\caption{Model of the MiniCLEAN detector installed in the water shield tank in the SNOLAB Cube Hall.}
\label{mc}
\end{center}
\end{figure}

The Inner Vessel is a stainless steel pressure vessel approximately 64'' in diameter.  Figure \ref{IV} shows the fully assembled Inner Vessel inside a softwall cleanroom underground.  The Inner Vessel was fabricated as a ASME Div. 1 Sec. VIII pressure vessel by Winchester Precision Technology in Winchester, NH.  The Inner Vessel is contained within a vacuum cryostat called the Outer Vessel, a stainless steel vessel approximately 104'' in diameter and 106'' high (see Figure \ref{OV}).  This Outer Vessel provides thermal insulation and containment for the Inner Vessel.  The Outer Vessel is positioned on a detector stand that allows for seismic isolation of the detector.  This entire assembly is located inside a water shield tank that is 18' in diameter and 25' tall.  A deck structure above the tank allows for access to the top of the tank for assembly of the experiment and ancillary systems such as the cryogenic and process systems, calibration system, electronics and data acquisition, and magnetic compensation coils.  The water shield tank contains a muon veto system composed of 48 additional photomultiplier tubes.  Figure \ref{mc} shows a model of the MiniCLEAN central detector and its shield tank layout.

The MiniCLEAN detector utilizes a number of subsystems.  The target is purified as a gas using a SAES getter to remove contaminants that affect the scintillation light.  A charcoal trap is then used to remove any radon from the gas before it enters the Inner Vessel.  To liquify the cryogen, two copper cold heads are attached to 3'' ports on the Inner Vessel and are cooled by a Cryomech Gifford McMahon cryorefrigerator capable of providing 325 W at 87K and 100 W at 25K.  These cold heads cool the gas in the Inner Vessel until it liquifies and fills the detector volume.  The 92 photomultiplier tubes signals are digitized using CAEN V1720 250 MHz digitizers and custom data acquisition software.  The water shield tank around the detector is instrumented with 48 Hamamatsu R1408 photomultiplier tubes to serve as a muon veto system \cite{veto}.  Around the shield tank are six magnetic compensation coils to cancel the earth's vertical magnetic field in order to improve the performance of the detector's photomultiplier tubes \cite{magcomp}.

\section{Current Status}
As of early 2014, the MiniCLEAN detector is nearing completion.  All 92 optical modules have been installed in the Inner Vessel and the PMTs are undergoing testing with the full electronics and DAQ system.  Prior to moving the Inner Vessel into the Outer Vessel, data are being acquired from a fill of argon gas using the gas purification system.  Once the Inner Vessel is placed in the Outer Vessel all final connections will be made and the muon veto system will be deployed into the water shield tank before filling the tank with water.  Cooling of the detector with cold gas and liquification of the argon is scheduled to start by Summer 2014.  After a brief run to determine initial backgrounds the detector will begin a long run to attain data to test the pulse shape discrimination capability of liquid argon.

\section{Utility of $^{39}$Ar Spike}
The internal background in MiniCLEAN, dominated by $^{39}$Ar beta decay, is mitigated using pulse shape discrimination (PSD) made possible by the scintillation timing difference described above. A nucleus scattered by a WIMP produces far fewer triplet states than the electron from $^{39}$Ar beta decay, and hence has more ``prompt'' than ``late'' light.  In order to demonstrate the effectiveness of the pulse-shape discrimination capability of liquid argon, MiniCLEAN will deploy a ``spike'' of $^{39}$Ar with an activity level approximately ten times higher than natural argon.  Such a spike will improve existing measurements of PSD rejection by over two orders of magnitude. Should we have a set of WIMP candidates in a future argon detector, a spike of $^{39}$Ar can also be deployed in that detector to directly determine if $^{39}$Ar leakage is the cause. The intrinsically fast timing of the scintillation light in a single-phase detector makes the use of such spikes possible, because pile-up of events is a small effect.  In October 2013, approximately 1.7 $\mu$Ci of $^{39}$Ar was extracted from a potassium salt (KCl) target that had been irradiated at TRIUMF. 

\section{Utility of Liquid Neon Run for Dark Matter}
\begin{figure}[htbp]
\begin{center}
\includegraphics[width=3in]{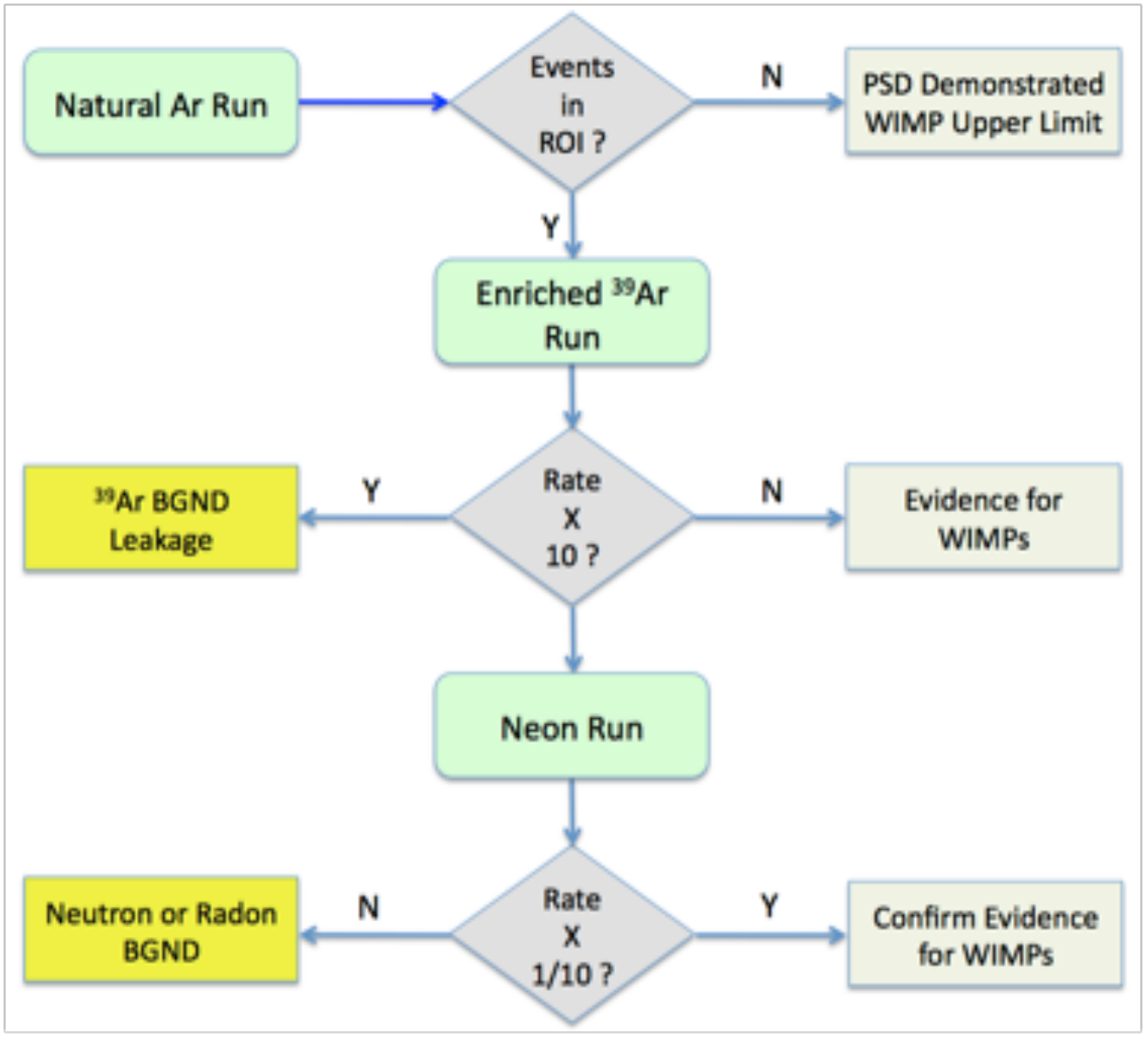}
\caption{Flow diagram showing the approach to verify a positive signal in a liquid argon/neon detector.}
\label{neon}
\end{center}
\end{figure}

Exploiting liquid neon (LNe) as a target has the advantage that there are no internal backgrounds like $^{39}$Ar, and thus the requirements for pulse shape discrimination are far lower. Building a detector that can use either LNe or LAr targets would allow one to use the difference in WIMP cross section as an additional way of verifying any putative signal. With LNe in place of LAr, the number of detected WIMP events should drop by nearly a factor of 10, much like a ``beam off'' measurement, while the number of neutrons that can cause background recoils will remain the same, apart from predictable differences in detector response between the two targets. Together with a $^{39}$Ar spike, a LNe deployment provides us with the flow diagram of signal verification shown in Figure \ref{neon}. The ability to swap targets and to test directly a possible signal is a powerful and unique capability of the single-phase approach. 

\section{Analysis Techniques}
\begin{figure}[htbp]
\begin{center}
\includegraphics[width=2.9in]{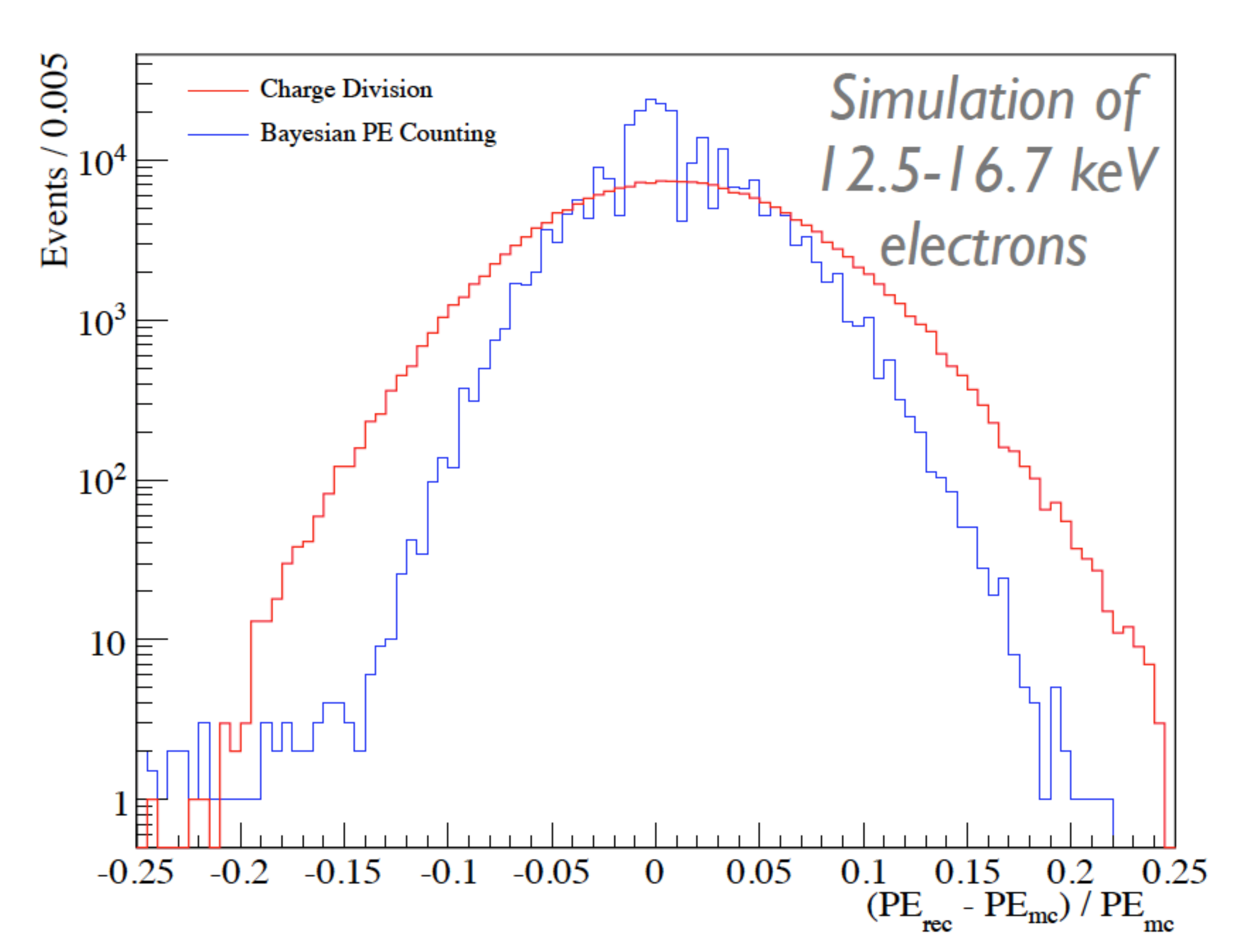}
\includegraphics[width=2.9in]{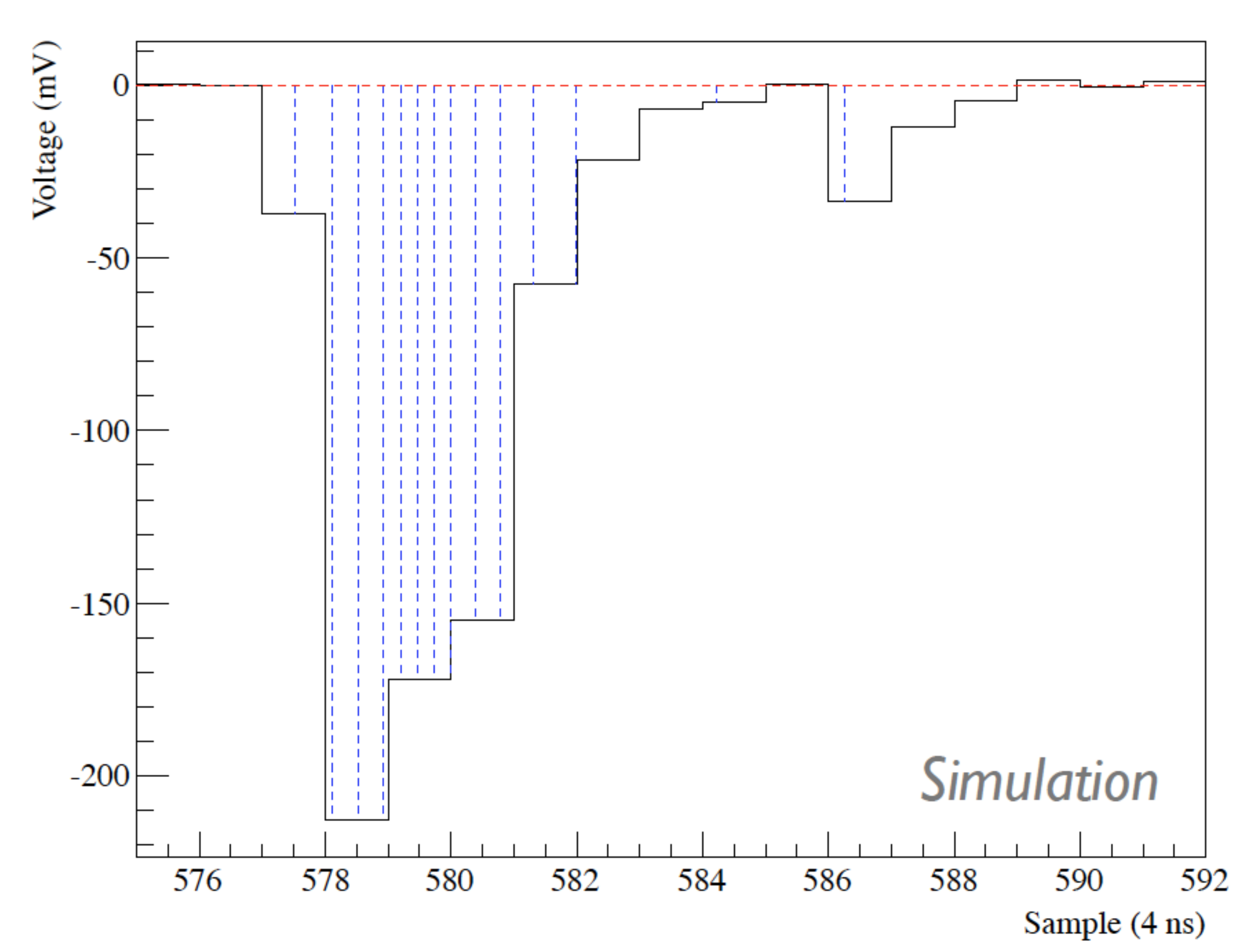}
\caption{The difference in photoelectron number in an event between the reconstructed number from both charge division and Bayesian counting and the actual number from simulated events (left).  A simulated waveform with predicted photoelectrons and timing from the Bayesian photoelectron method (right).}
\label{bay}
\end{center}
\end{figure}

\begin{figure}[htbp]
\begin{center}
\includegraphics[width=4in]{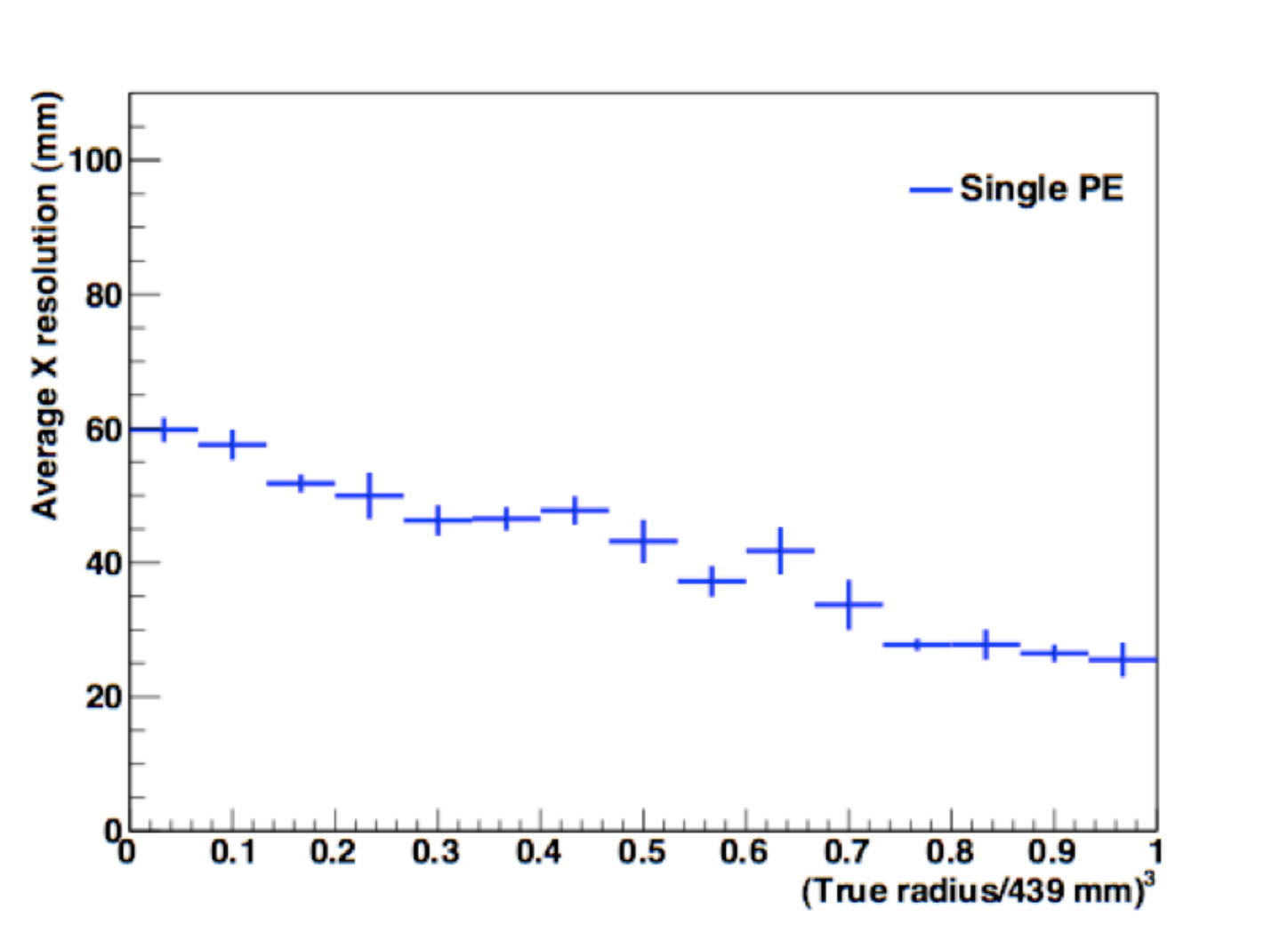}
\caption{Average reconstructed position resolution (in the x coordinate) vs. the true radius in the detector for simulated 20 keV events.}
\label{res}
\end{center}
\end{figure}

\begin{figure}[htbp]
\begin{center}
\includegraphics[width=2.73in]{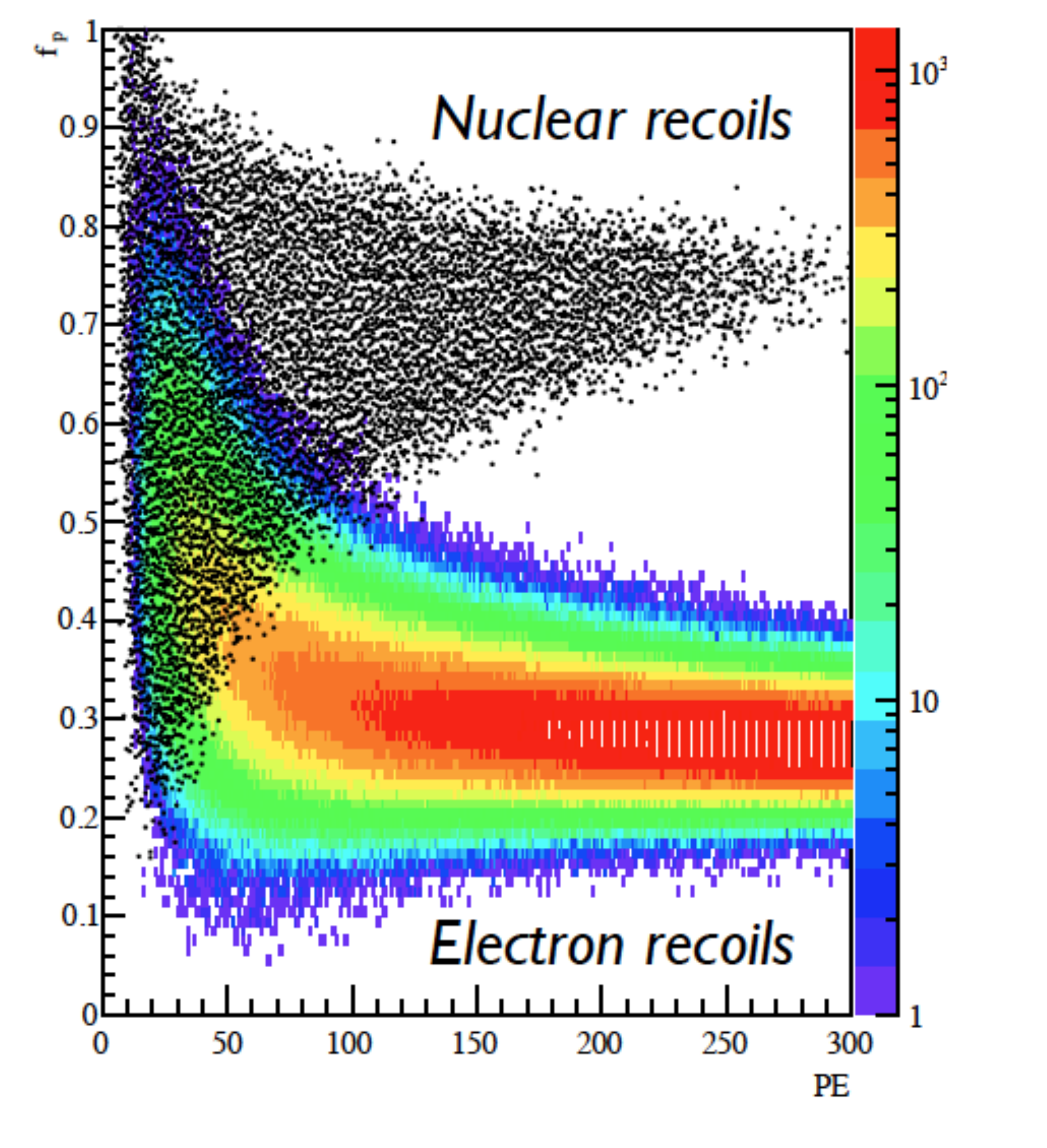}
\includegraphics[width=2.6in]{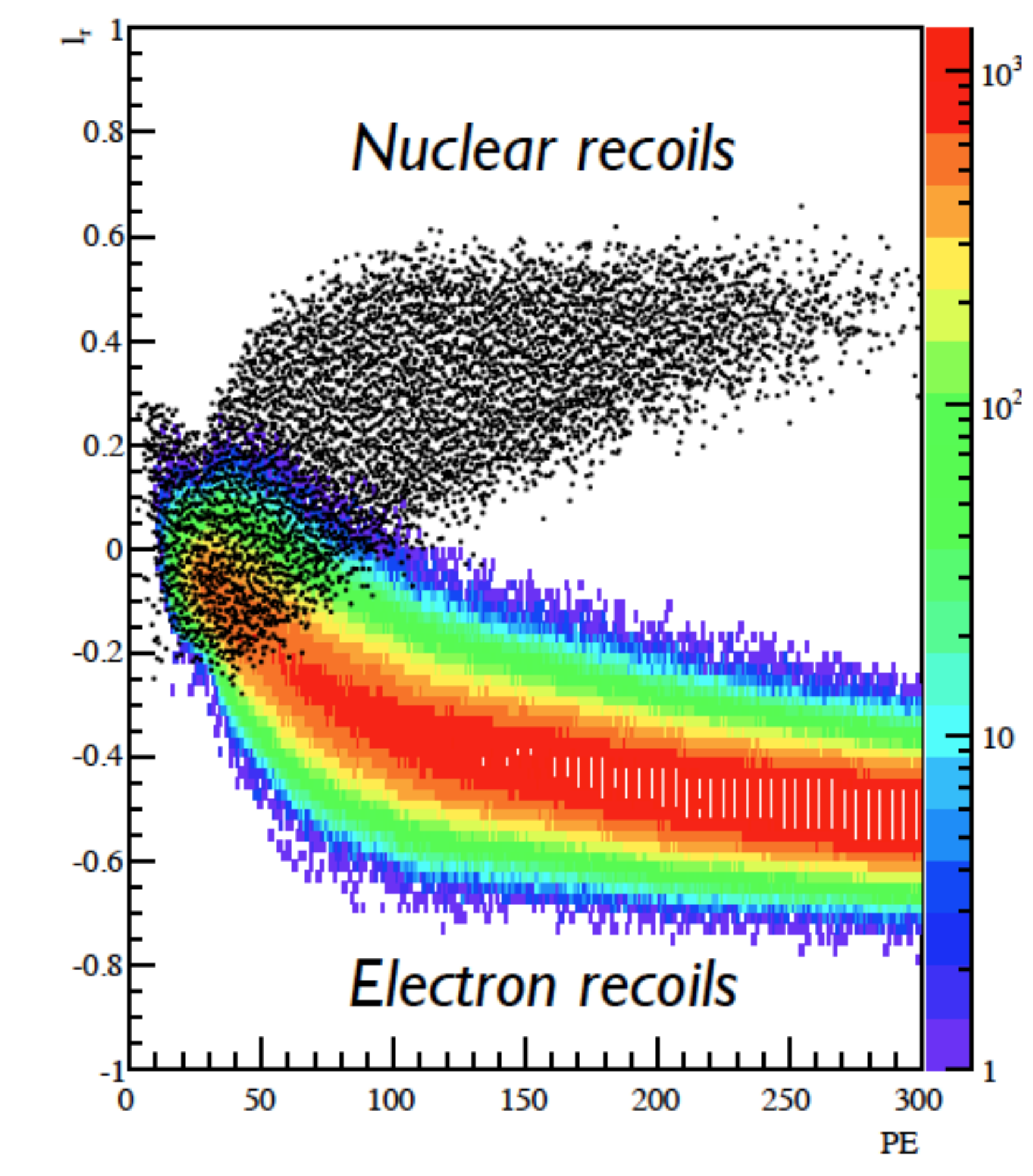}
\caption{The Fprompt distribution for simulated WIMPs and $^{39}$Ar recoil events as a function of energy in photoelectrons (left). The Lrecoil distribution for simulated WIMPs and $^{39}$Ar recoil events as a function of energy in photoelectrons (right) shows signifiant improvement in discrimination capability at lower energy.  For reference, MiniCLEAN expects approximately 6 photoelectrons per keVee for its light yield.}
\label{fprompt}
\end{center}
\end{figure}

The MiniCLEAN detector's response can be simulated using a Monte Carlo framework based on GEANT4.  Events can be reconstructed by energy, position and type.  In order to accurately reconstruct the total energy of the event it is necessary to determine the number of photoelectrons in the recorded waveform.  This can be done by simply integrating the waveform and dividing by the average charge of a single photoelectron or one can use a Bayesian analysis technique to determine the probability, timing and number of photoelectrons in the waveform.  Figure \ref{bay} compares the two approaches using simulated data and the best fit for a simulated waveform from the Bayesian technique.  Using the distribution of the light across the detector, the event can be reconstructed within the detector by using a self-tuning maximum likelihood fitter.  Figure \ref{res} shows the average position reconstruction resolution in the x-coordinate as a function of the true position of the event in the detector for simulated 20 keV events.  The resolution is significantly better at the edge of the detector which enables better rejection of background events and the ability to determine a clean fiducial volume at the center of the detector for a WIMP search.  The discrimination between electron recoil events and nuclear recoil events utilizes the difference in the triplet and singlet decay times.  A simple ratio of the amount of prompt light (first 100 ns) to the total light in the event (called Fprompt) can effectively discriminate at high energies (greater than 30 keVee).  Using a likelihood ratio (called Lrecoil) based on the timing of the photoelectrons from the Bayesian technique discussed above significantly improves the discrimination capability at lower energies.  Figure \ref{fprompt} shows the distribution of simulated nuclear recoil and electron recoil events for the two approaches as a function of energy in photoelectrons.  By using these three variables (position, energy, and type), backgrounds can be reduced within the central fiducial volume and sensitivity to potential WIMP events can be maximized.

\section{Conclusions}
The MiniCLEAN detector is nearing completion at the SNOLAB underground laboratory in Sudbury, Canada.  This unique single-phase noble liquid detector will demonstrate the pulse shape discrimination capability of liquid argon using a spike of $^{39}$Ar.  The detector design also allows for the use of liquid neon as a target in order to verify that any signal from potential WIMP events are not from detector backgrounds.  MiniCLEAN is expected to begin operations later in 2014 and serve as a demonstration for a future detector with a target mass of 50 to 100 tonnes.

\section{Acknowledgments}
The MiniCLEAN experiment is supported by the LANL LDRD program, the Department of Energy Office of Science, the National Science Foundation, the European Research Council, and the Packard Foundation.  We gratefully acknowledge the support and assistance of SNOLAB.  Contributions by staff of NIST, an agency of the US Government,
are not subject to copyright in the United States.





\bibliographystyle{elsarticle-num}
\bibliography{<your-bib-database>}

\begin{thebibliography}{00}


\bibitem{BH}
M.~G. Boulay and A. Hime, Astropart. Phys. \emph{25,} 179 (2006).

\bibitem{clean}
D.~N. McKinsey and K.~J. Coakley, Astropart. Phys. \emph{22,} 355 (2005).

\bibitem{Lipp}
W.~H. Lippincott, et al., Phys. Rev. C \emph{78,} 035801 (2008).

\bibitem{Nik}
J.~A. Nikkel, R. Hasty, W.~H. Lippincott, and D.~N. McKinsey, Astropart. Phys. \emph{29,} 161 (2008).

\bibitem{DEAP1}
M.~G. Boulay, et al., arXiv:0904.2930 (2009).

\bibitem{argon}
M.~G. Boulay, A. Hime, and J. Lidgard, arXiv:nucl-ex/0410025 (2004).

\bibitem{pmt}
J.~A. Nikkel, W.~H. Lippincott, and D.~N. McKinsey, J. Inst. \emph{2,} P11004 (2007).

\bibitem{TPB}
V.~M. Gehman, et al., arXiv: 1104.3259 (2011).

\bibitem{acrylic}
M. Bodmer, N. Phan, M. Gold, J.~A.~J. Matthews, and K. Rielage, J. Inst. \emph{9}, P02002 (2014).

\bibitem{veto}
R. Abruzzio, B. Buck, S. Jaditz, J. Kelsey, J. Monroe, and K. Palladino, arXiv:1403.1549 (2014).

\bibitem{magcomp}
M. Bodmer, F. Giuliani, M. Gold, A Christou, and M. Batygov, Nucl. Instr. Meth. \emph{A697}, 99 (2013).



\end{thebibliography}



\end{document}